\documentclass[aip,nofootinbib,letterpaper,reprint]{revtex4-2} 
\usepackage{amsmath,amsthm,amssymb}
\usepackage{bbm}
\usepackage{xcolor}
\usepackage{graphicx}
\usepackage{mathrsfs} 
\usepackage{url}
\usepackage[colorlinks=true,breaklinks=true,allcolors=blue]{hyperref}
\usepackage{braket}

\newcommand\RR{{\mathbbm{R}}}

\newcommand{\mvec}[1]{\boldsymbol #1}

\definecolor{michael}{rgb}{0,.8,.5}

\begin{document} 

\title{Long-range systems, (non)extensivity, and the rescaling of energies} 

\author{Michael Kastner} 
\affiliation{Institute of Theoretical Physics, Stellenbosch University, Stellenbosch 7600, South Africa\\ Hanse-Wissenschaftskolleg, Lehmkuhlenbusch 4, 27753 Delmenhorst, Germany}

\date{\today}
 
\begin{abstract}
Systems with long-range interactions have seen a surge of interest in the past decades. In the wake of this surge, the use of a system size-dependent rescaling, sometimes termed ``Kac prescription,'' of the long-range pair potential has seen widespread use. An {\em ad hoc}\/ modification of the Hamiltonian may make the energy extensive, but its physical justification and implications are a frequent source of confusion and misinterpretation. After all, in real physical $N$-body systems, the pair interaction strength does not scale with the number $N$ of constituents. This article presents, at an introductory level, scaling arguments that provide a clear physical interpretation of the ``Kac prescription'' for finite systems as well as in the thermodynamic limit.
\end{abstract}


\maketitle 


\section{Introduction}
\label{s:Introduction}
A quantity is called extensive if it scales linearly with the number $N$ of the constituents of a many-particle system. Extensivity plays an important role in physical theories that apply to many-particle systems, like thermodynamics or statistical physics. Examples include quantities like the entropy $S$ and the energy $E$ of a system, which, in many cases, are found to scale for large $N$ like
\begin{equation}
E\sim Ne,\qquad S\sim Ns,
\end{equation}
where $e$ and $s$ are $N$-independent quantities. However, there are physical situations in which extensivity is violated. An example is a gas of particles of mass $m$, interacting via a gravitational pair potential
\begin{equation}\label{e:Gpot}
U=-\sum_{i<j}\frac{Gm^2}{|\mvec{r}_i-\mvec{r}_j|},
\end{equation}
where $\mvec{r}_i$ denotes the position vector of particle $i$, and $G$ is the gravitational constant. Confining such a gas in a cubic box of volume $V$ with periodic boundary conditions and assuming the particles to be distributed uniformly in space with a density $N/V$, one can estimate the total potential energy to scale like \cite{CampaDauxoisFanelliRuffo}
\begin{equation}\label{e:U53}
U\sim N^{5/3}\times\,\text{const.}
\end{equation}
In this case the energy is over-extensive, scaling like $N^q$ with an exponent $q>1$. The over-extensivity is a consequence of the gravitational potential \eqref{e:Gpot} being long-range: In $D$ spatial dimensions, the pair-potential decays slower than $1/r^D$ with the distance $r=|\mvec{r}|$ between two particles.

Much of the formalism of thermodynamics and statistical physics relies on the extensivity of certain quantities, and in particular on the extensivity of the energy. So much so in fact that, when dealing with long-range interacting systems, it has become customary to {\em enforce}\/ the energy to be extensive, a procedure that has become known under the name of Kac prescription in the statistical physics context;\cite{CampaDauxoisFanelliRuffo,Defenu_etalRMP23} see Appendix~\ref{s:history} for historical remarks. As an example, consider $N$ classical Ising spins $s_i\in\{-1,+1\}$ with all-to-all interactions, i.e., each spin interacts with every other spin with equal strength. The energy function of such a spin system is
\begin{equation}\label{e:CWIsing}
H_h=-J\sum_{i<j}s_is_j-h\sum_is_i,
\end{equation}
where $J$ is the spin-spin coupling and $h$ the magnetic field strength. The sums in \eqref{e:CWIsing} can be bounded by
\begin{equation}\label{e:sum_bounds}
\Bigl|\sum_is_i\Bigr|\leq N,\qquad \Bigl|\sum_{i<j}s_is_j\Bigr|\leq\frac{N(N-1)}{2},
\end{equation}
and the bounds are saturated for the fully magnetized state where $s_i=+1$ (or $s_i=-1$) for all $i=1,\dotsc,N$. The second inequality in \eqref{e:sum_bounds}, when saturated, attests that the energy is over-extensive, scaling like $N^2$ for large $N$. Extensivity of the energy function \eqref{e:CWIsing} can be enforced by an $N$-dependent rescaling of the coupling constant $J$,
\begin{equation}\label{e:CWIsingRescaled}
\widetilde{H}_h=-\frac{J}{N}\sum_{i<j}s_is_j-h\sum_is_i,
\end{equation}
where the tilde distinguishes the rescaled Hamiltonian from its original (unscaled) counterpart.

While $\widetilde{H}_h$ has been made extensive, it remains {\em nonadditive}: Unlike in short-range interacting systems, when cutting the lattice in two halves, the energies of the two parts of a long-range model do not add up, in the large system limit, to the energy of the total system. Nonadditivity is at the heart of many of the physical peculiarities of long-range systems, and it is reassuring that this property, which we may want to study, remains intact upon rescaling of $H_h$.

For a lattice model with long-range interactions that are not all-to-all, different powers of $N$ may be needed for the rescaling in Eq.~\eqref{e:CWIsingRescaled}. Continuum models, like the gravitational potential \eqref{e:Gpot}, again differ in the prefactor needed to enforce extensivity. These different rescaling schemes will be introduced and discussed in Secs.~\ref{s:lattice} and \ref{s:continuum}. While enforcing extensivity is a relatively simple concept, the physical ``meaning'' of such an {\em ad hoc}\/ modification of the energy function is not only a matter of debate,\cite{Lorenzetti25} but also a frequent source of confusion and misinterpretation. After all, the pair interaction strength between the constituents of a many-body system does {\em not}\/ scale with the number $N$ of constituents in most models of physical interest. Still, results obtained from a rescaled energy function like \eqref{e:CWIsingRescaled}, be it for finite systems or in the thermodynamic limit $N\to\infty$, can be given a clear physical interpretation, to which the reader is introduced in Secs.~\ref{s:finiteN}--\ref{s:dynamics}.

The present article aims at providing conceptual clarity and a lucid interpretation of existing and forthcoming results in the vibrant research areas of classical and quantum long-range interacting systems. The selection of material, and also the level of presentation, is pitched such that it should be useful for instructors of physics, graduate students, and more senior newcomers to the field alike.


\section{Long-range lattice models}
\label{s:lattice}
Going beyond the all-to-all interactions of \eqref{e:CWIsingRescaled} where the interaction strength between two sites $i$ and $j$ is independent of their distance $|i-j|$,\footnote{In two- or higher-dimensional lattices, site indices become vector-valued, with the Euclidean distance between these vector indices usually being the relevant distance in physical Hamiltonians.} we consider here a more general class of long-range lattice models, characterized by coupling strengths proportional to some inverse power of the distance $|i-j|$, as in
\begin{equation}\label{e:LRIsing}
\widetilde{H}_h=-J\mathcal{N}_N\sum_{i<j}\frac{s_is_j}{|i-j|^\alpha}-h\sum_is_i
\end{equation}
with $\alpha\geq0$. While a classical Ising model is used as an example, the following reasoning extends straightforwardly to general spin models or lattice gases, be they classical or quantum. On a regular lattice consisting of $N$ sites in $D$ spatial dimensions, a short calculation shows that the energy function \eqref{e:LRIsing} can be made extensive by choosing an $N$-dependent prefactor
\begin{equation}\label{e:Prefactor}
\mathcal{N}_N\propto1\Bigl/\sum_{j\neq0}\frac{1}{|j|^\alpha},
\end{equation}
where the summation is over all $N$ lattice sites except the one at the origin. For large $N$, this sum can be estimated by an integral approximation, yielding\cite{AnteneodoTsallis98}
\begin{equation}\label{e:curlyN}
\mathcal{N}_N\sim
\begin{cases}
N^{\alpha/D-1} &\text{for $0\leqslant\alpha<D$},\\
1/\ln N &\text{for $\alpha=D$},\\
\text{const.} &\text{for $\alpha>D$}.
\end{cases}
\end{equation}
This result can be interpreted as follows: For $\alpha>D$, the interaction strength decays sufficiently fast with the distance so that the double sum in Eq.~\eqref{e:LRIsing}, even though consisting of $N^2$ terms, gives a contribution of order $N$. Accordingly, no $N$-dependent prefactor is needed to make $\widetilde{H}_h$ extensive.
For $\alpha\leq D$ on the other side, the nonnegligible coupling to distant sites makes a rescaling necessary to enforce extensivity. For $\alpha=0$, the $1/N$ scaling of the all-to-all interactions \eqref{e:CWIsingRescaled} is recovered. From this discussion it becomes clear that, in long-range lattice models, it is the long-distance behavior of the interaction potential that may cause nonextensive energies.


\section{Continuum models}
\label{s:continuum}
In continuum models, nonextensivity may also be caused by the short-distance behavior of the pair potential when the system is in a collapsed phase. This occurs when considering a gas of particles with attractive interactions, like point masses in a regularized gravitational potential,
\begin{equation}\label{e:GpotReg}
H=\sum_i\frac{\mvec{p}_i^2}{2m}-\sum_{i<j}\frac{Gm^2}{\sqrt{|\mvec{r}_i-\mvec{r}_j|^2+a^2}},
\end{equation}
where $a>0$ is a short-distance cutoff.\cite{Plummer11} Such a cutoff is necessary in order to have a finite energy in the collapsed phase where most particles accumulate in one region of space. On physical grounds, the cutoff can be seen as a way of incorporating quantum effects (like the Pauli exclusion principle) that become relevant on short length scales. In the collapsed phase, when a significant fraction of the $N$ particles forms a cluster, $|\mvec{r}_i-\mvec{r}_j|\approx0$ holds approximately and hence the potential energy term in \eqref{e:GpotReg} scales quadratically with $N$,
\begin{equation}\label{e:continuumscaling}
-\sum_{i<j}\frac{Gm^2}{a}\sim N^2.
\end{equation}
For the Hamiltonian to be extensive, energies must grow at most linearly with $N$ for any type of particle configuration. For large system sizes, the $N^2$-scaling due to collapsing particles in Eq.~\eqref{e:continuumscaling} is dominant compared to the $N^{5/3}$ scaling in Eq.~\eqref{e:U53} caused by the long-range tails of the gravitational potential. Accordingly, a rescaling factor of $1/N$ is appropriate to enforce extensivity of the potential energy.
Since that factor is determined by the short-distance properties of the pair potential, and not by its long-distance decay, a general power law potential
\begin{equation}\label{e:GpotAlpha}
\widetilde{U}=-\frac{1}{N}\sum_{i<j}\frac{Gm^2}{\sqrt{|\mvec{r}_i-\mvec{r}_j|^2 + a^2}^\alpha},
\end{equation}
requires the same $1/N$ rescaling to enforce extensivity, independently of the value of the long-range exponent $\alpha$.

Extensivity in statistical physics and thermodynamics is a concept that refers to large systems. Accordingly, the thermodynamic limit of infinite system size will play an important role in Secs.~\ref{s:why}--\ref{s:dynamics}. However, more accurately, there is no such thing as {\em the} thermodynamic limit but, depending on the type of system and the physical context, different limiting procedures can be performed in which the system size is sent to infinity. Ambiguities already arise for lattice systems where the limit is taken over a sequence of lattices of increasing size, and the shape of these lattices may matter, in particular their boundary-to-volume ratio.\cite{Ruelle} The multitude of possible limits, and the ensuing mathematical complications, become even more involved for continuum models. 
I will avoid these complications in what follows by resorting to lattice models, tacitly assuming well-behaved shapes of lattices (e.g., hypercubes). The key concepts of the paper carry over to continuum models, as is briefly discussed at the end of Sec.~\ref{s:interpretation}.


\section{Rescaling in finite systems}
\label{s:finiteN}
In the preceding Secs.~\ref{s:Introduction}--\ref{s:continuum}, possible origins of nonextensive terms were discussed and suitable $N$-dependent prefactors that enforce extensivity in the presence of such terms were introduced. The latter is clearly an {\em ad hoc}\/ modification, and it must seem questionable whether it is permissible to tamper with the system of interest in this way. After all, in most cases the interaction potential of a physical system is given by Nature, and our goal is to compute properties of this system, not its modified variant.\footnote{In a few exceptional cases, it may be possible to experimentally vary the pair coupling strength in a controlled manner when increasing the system size. Examples include quantum simulators based on trapped ultracold atoms or ions where the experimenter has a high degree of control over the coupling constants, which in some cases can be varied over several orders of magnitude.\cite{BlochDalibardNascimbene12,BlattRoos12}} In this and the subsequent sections the following questions are explored:
\begin{enumerate}
\renewcommand{\labelenumi}{(\alph{enumi})}
\item Is it justifiable to introduce $N$-dependent scaling factors in the Hamiltonian to make it extensive?
\item When is it useful and advisable to do so?
\end{enumerate}
These questions are discussed for the case of finite systems in the present section, and for systems in the thermodynamic limit in Secs.~\ref{s:why}--\ref{s:dynamics}.

For concreteness, we consider the spin Hamiltonian \eqref{e:LRIsing} with prefactor \eqref{e:curlyN}. It can be considered justified to introduce the $N$-dependent prefactor $\mathcal{N}_N$ in the Hamiltonian as long as one can translate results obtained for the rescaled Hamiltonian $\widetilde{H}_h$ back to statements about the unscaled Hamiltonian
\begin{equation}
H_h=-J\sum_{i<j}\frac{s_is_j}{|i-j|^\alpha}-h\sum_is_i.
\end{equation}
In the context of equilibrium statistical physics, a central quantity of interest is the canonical partition function $Z_H$ of the Hamiltonian $H_h$. Writing out the partition functions for the unscaled Hamiltonian $H_h$ and its scaled counterpart $\widetilde{H}_h$, it is straightforward to read off that
\begin{multline}\label{e:ZZ}
Z_{\widetilde{H}}(\beta,h) = \sum_x \exp\left[-\beta \widetilde{H}_h(x)\right]\\
= \sum_x \exp\left[-\beta\mathcal{N}_N H_{h/\mathcal{N}_N}(x)\right] = Z_{H}(\beta\mathcal{N}_N,h/\mathcal{N}_N).
\end{multline}
Here $\beta=1/(k_\text{B}T)$ is the inverse temperature and $k_\text{B}$ denotes the Boltzmann constant. The notation used here is that of a classical model with discrete degrees of freedom, where $x$ denotes a many-particle microstate of the system.\footnote{In the case of the classical spin Hamiltonian \eqref{e:CWIsing}, each $x\in\{-1,+1\}^{\times N}$ is a specific string of spin orientations, e.g., $\{++-+\cdots\}$, and the summation in Eq.\ \eqref{e:fN} is over all $2^N$ different such strings. When dealing with classical continuum models like in Eq.\ \eqref{e:GpotReg}, the sum is replaced by a phase space integral. For quantum mechanical models, like the quantum spin model \eqref{e:LRQuantumIsing} discussed in Sec.~\ref{s:dynamics}, a trace over the Hilbert space replaces the sum over microstates in the definition of the canonical free energy density.}  Equation \eqref{e:ZZ} illustrates that, by computing the canonical partition function $Z_{\widetilde{H}}$ with respect to the rescaled Hamiltonian $\widetilde{H}_h$ at certain values of $\beta$ and $h$, one gains information also about the partition function of the unscaled Hamiltonian $H_h$ at appropriately rescaled values of $\beta$ and $h$. Hence, on a mathematical level it is justified to use, if one so wishes, the rescaled Hamiltonian $\widetilde{H}_h$ for doing calculations, because results can be translated into statements about the original Hamiltonian $H_h$. This reasoning is not restricted to partition functions, but applies to other quantities as well; see Sec.~\ref{s:dynamics} for another example.

While the above explanations demonstrate that the rescaling of overextensive terms in the Hamiltonian is permissible, we have yet to see a good reason for doing so. After all, since the results are equivalent in the sense that they can be translated into each other, little appears to be gained. This is indeed often true for finite systems. In the thermodynamic limit of infinite system size, however, the scaling factors in Eq.\ \eqref{e:ZZ} become singular, the transformation between $Z_H$ and $Z_{\widetilde{H}}$ breaks down, and the results cease to be equivalent. We will explore this situation in the following Secs.~\ref{s:why} and \ref{s:interpretation}.


\section{Rescaling in the thermodynamic limit}
\label{s:why}
A frequently cited reason for the necessity of an extensive energy is ``to make statistical physical quantities well-defined in the thermodynamic limit.'' A central quantity in statistical physics that can be used to illustrate this statement is the canonical free energy density
\begin{equation}\label{e:fN}
f_N(\beta,h)=-\frac{1}{N\beta}\ln Z_H(\beta,h).
\end{equation}
This quantity is designed to have a well-defined (and nontrivial) thermodynamic limit $N\to\infty$ precisely when $H_h$ is extensive: The partition function $Z_H$ then consists of exponentials of the extensive Hamiltonian; taking the logarithm of that quantity takes us back to an extensive quantity $\propto N$; which is then divided by $N$, making it $\mathcal{O}(1)$, hence resulting in a well-defined thermodynamic limit. However, this does not necessarily imply that making the Hamiltonian extensive by introducing the scaling factor $\mathcal{N}$ is the only way to deal with this situation. Instead, one could redefine $f_N$ with a prefactor $\propto \mathcal{N}_N/N$ (instead of $1/N$), which has a well-defined thermodynamic limit by construction, independently of whether the Hamiltonian is extensive or not.

So why is this latter strategy not usually followed and why does one instead decide to tamper with the Hamiltonian? The reason is that much of the interesting physics, and in particular the occurrence of phase transitions, has its origin in the presence of {\em competing}\/ terms in the Hamiltonian and, as will be explained in the following, modifying the canonical free energy density may eliminate this competition of terms, thereby removing the interesting physics. This is because, while a part of the Hamiltonian may scale overextensively and need to be rescaled, other parts may be extensive from the outset. In our example of the all-to-all Ising spin model \eqref{e:CWIsing}, only the first sum on the right-hand side scales like $N^2$, whereas the second sum grows linearly with $N$. 
If we now were to deal with the nonextensivity by modifying the prefactor of the canonical free energy density \eqref{e:fN}, the extensive contribution from the second term of the Hamiltonian \eqref{e:CWIsing} would vanish in the thermodynamic limit,
\begin{flalign}
\tilde{f}(\beta,h) &= -\lim_{N\to\infty}\frac{1}{N^2\beta}\label{e:fN_mod}\\
&\quad\times\ln\sum_x \exp\biggl(\beta J\sum_{i<j}s_i s_j + \beta h\sum_i s_i\biggr)\nonumber\\
&= -\lim_{N\to\infty}\frac{1}{N^2\beta}\ln\sum_x \exp\biggl(\beta J\sum_{i<j}s_i s_j\biggr) = \tilde{f}(\beta,0)\nonumber
\end{flalign}
for arbitrary $h$. All physical effects originating from the presence of the field term in large but finite systems would be lost in the thermodynamic limit.

Similarly one can show that entropic effects, originating from the product structure of many-body phase spaces or Hilbert spaces and the resulting exponential growth (in $N$) of the microcanonical density of states, are eliminated when using a nonextensive Hamiltonian together with the modified definition of the canonical free energy in Eq.\ \eqref{e:fN_mod}. To retain a competition between entropic and energetic effects when taking the thermodynamic limit, one needs to make sure that the Hamiltonian scales with $N$ in the same way as the entropy does: Both should be extensive.


\section{Rescaling and the real world}
\label{s:interpretation}
It may be tempting to dismiss the arguments brought forward in Sec.~\ref{s:why} by declaring that, well, if it turns out that in the thermodynamic limit there is no competition between, say, energy and entropy in the long-range Ising model \eqref{e:LRIsing}, then we must accept this outcome. But this falls short of the mark. To understand why, it is useful to recapitulate the role of the thermodynamic limit $N\to\infty$ in physics. Real physical systems are finite, and the thermodynamic limit is a simplifying idealization, which may be useful as long as no oversimplification occurs. 
Mathematically, limit theorems may allow one to obtain analytic results in the thermodynamic limit, whereas such results for large but finite systems are often out of reach. Also, in the thermodynamic limit a phase transition has a clear definition as a nonanalyticity in some thermodynamic function, whereas the smeared out cross-over observed in a finite system is more difficult to pinpoint. An experimentalist, on the other side, will make implicit use of the concept of the thermodynamic limit when telling you that they have measured, say, the specific heat of copper. Since they are not mentioning the size and shape of the copper block, they are implicitly stating that the sample used in the measurement was large enough for surface effects not to play a role and that the bulk (thermodynamic limit) value of the specific heat was measured. 

\begin{figure}\center
\includegraphics[width=0.8\linewidth]{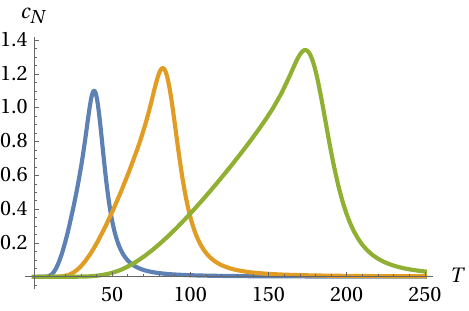}
\includegraphics[width=0.8\linewidth]{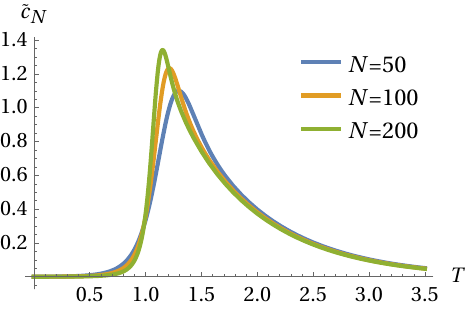}
\caption{\label{f:SpecificHeat}
(Color online) The specific heat as a function of the temperature $T$ for the all-to-all Ising model for different system sizes $N$. Top: Specific heat \eqref{e:SpecificHeat} for the Hamiltonian \eqref{e:CWIsingUnscaled_h0} without an $N$-dependent prefactor. The posititon of the peak shifts to larger $T$ with increasing $N$. Bottom: As in the top panel, but for the scaled Hamiltonian \eqref{e:CWIsingRescaled_h0}. The larger $N$, the higher and sharper the peaks become.
}
\end{figure}

Assume now that we decide to use the all-to-all Ising Hamiltonian \eqref{e:CWIsing} {\em without}\/ an $N$-dependent prefactor,
\begin{equation}\label{e:CWIsingUnscaled_h0}
H_{\text{a2a}}=-\sum_{i<j}s_is_j,
\end{equation}
where, to keep the argument simple, we have set the interaction strength $J$ to unity and the magnetic field $h$ to zero. The canonical free energy density of this model can be written in the form \cite{KochmanskiPaszkiewiczWolski13}
\begin{equation}\label{e:fN_CurieWeiss}
f_N(\beta)=-\frac{1}{N\beta}\ln\left[\frac{2^N e^{-\beta}}{\sqrt{2\pi}} \int_\RR d\xi\, e^{-\xi^2/2}\cosh\left(\xi\sqrt{\beta}\right)^N\right],
\end{equation}
from which the specific heat
\begin{equation}\label{e:SpecificHeat}
c_N(T):=-T\frac{\partial^2 f_N(1/T)}{\partial T^2}
\end{equation}
can be computed as a function of the temperature $T=1/\beta$. (I use units where the Boltzmann constant $k_B=1$.) Figure~\ref{f:SpecificHeat} (top) shows plots of this quantity for different system sizes $N$, obtained by numerical evaluation of the integral in Eq.\ \eqref{e:fN_CurieWeiss}. The specific heat shows prominent peaks, which can be interpreted as finite-system precursors of a phase transition and are arguably the physically most interesting features in the plot. However, the plot also shows that, with increasing system size $N$, the peaks will shift to higher and higher temperatures, and will eventually disappear toward ``infinite temperature.'' Because of that shift to infinity, any sign of the peak and the phase transition will be absent from the thermodynamic limit result of the specific heat. The purpose of introducing the thermodynamic limit, however, was to obtain an idealized characterization of the behavior of large but finite systems, whose specific heat does show a peak that gets increasingly sharp for large $N$. So we must admit that we failed to achieve our goal. This does not mean that the thermodynamic limit gave us an incorrect result, but that the result is valid in a regime that we may consider as not particularly relevant or interesting. We will get back to this interpretation at the end of this section.

Along similar lines, one can calculate the specific heat $\tilde{c}_N$ for the all-to-all Ising Hamiltonian with an $N$-dependent prefactor included,
\begin{equation}\label{e:CWIsingRescaled_h0}
\tilde{H}_{\text{a2a}}=-\frac{1}{N}\sum_{i<j}s_is_j.
\end{equation}
The results are shown in Fig.~\ref{f:SpecificHeat} (bottom). Here, unlike in the case of the unscaled Hamiltonian \eqref{e:CWIsingUnscaled_h0}, the peaks of $\tilde{c}_N$ for different $N$ lie on top of each other and converge toward a sharp peak at $T=1$ in the thermodynamic limit. The thermodynamic limit yields an idealized characterization of the behavior of large but finite systems at temperatures corresponding to the peaks of the specific heat, which is the parameter region of interest.

\begin{figure}\center
\includegraphics[width=0.95\linewidth]{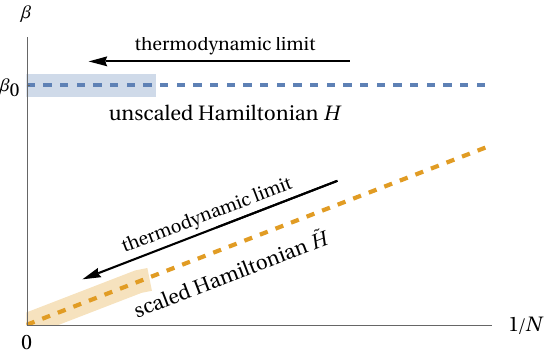}
\caption{\label{f:Limits}
(Color online) Illustration of the parameter regimes for which the thermodynamic limit gives an idealized description when using $\tilde{H}$ or $H$, respectively. Taking the thermodynamic limit of, say, the specific heat of the unscaled Hamiltonian $H$, one can think of fixing the inverse temperature $\beta$ to some value $\beta_0$ and considering the limit of $c_N(1/\beta_0)$ for larger and larger systems sizes $N$. The idea behind taking the thermodynamic limit along the blue dashed (horizontal) line is to obtain an idealized description or approximation valid at $\beta=\beta_0$ for sufficiently large $N$, as indicated by the shaded area around that line. Alas, for the example shown in Fig.~\ref{f:SpecificHeat} (top) the peaks of $c_N$ move out to larger and larger temperatures, resulting in a trivial limit value $c_\infty(1/\beta_0)=0$ for any nonzero $\beta_0$. On the other hand, using the scaled Hamiltonian $\tilde{H}$ when performing the thermodynamic limit corresponds, according to Eq.\ \eqref{e:ZZ}, to considering the unscaled Hamiltonian $H$ at rescaled values $\tilde{\beta}=N\beta$ of the inverse temperature. This limiting procedure along a path of nonconstant $\beta$ is indicated by the yellow dashed (slanting) line, and the validity of the idealization by the shaded area around that line. Rescaling $\beta$ amounts to shifting the peaks of $c_N$ for different system sizes to lie on top of each other, as in Fig.~\ref{f:SpecificHeat} (bottom), resulting in a nontrivial thermodynamic limit along that latter path.
}
\end{figure}

Since we introduced the thermodynamic limit with the goal to obtain an idealized description of the physics of large but finite systems, the above example illustrates that this goal is achieved when using the rescaled Hamiltonian \eqref{e:CWIsingRescaled_h0}, but not with the unscaled Hamiltonian \eqref{e:CWIsingUnscaled_h0}. Moreover, Eq.~\eqref{e:ZZ} shows that quantities obtained with the unscaled Hamiltonian $H$ at inverse temperature $\beta$ correspond to quantities obtained with the rescaled Hamiltonian $\tilde{H}$ at an inverse temperature $\tilde{\beta}=N\beta$. This correspondence implies that performing the thermodynamic limit for $\tilde{H}$ gives an idealized description of large but finite systems with Hamiltonian $H$ that is valid in a different region of parameter space than that of the unscaled Hamiltonian; see Fig.~\ref{f:Limits} for an illustration. In short, the physically more interesting region gets into focus when using the rescaled, extensive Hamiltonian.

We used the all-to-all Ising Hamiltonian as an example and the specific heat as the quantity of interest, but the outlined ideas and principles are much more broadly applicable. For instance, the larger the number of particles in a gravitating gas, the higher is the temperature (or kinetic energy) required to stabilize the gas against a gravitational collapse. Making the Hamiltonian extensive by introducing an $N$-dependent prefactor in the gravitational potential, the transition temperature of the model gets rescaled and, under suitable further conditions, may show a well-defined and nontrivial behavior in a suitably defined thermodynamic limit. Similar considerations apply in the context of quantum phase transitions at zero temperatures; see Sec.~IV, and in particular Sec.~IV.D, of Ref.~\onlinecite{Defenu_etalRMP23}.



\section{Dynamics and nonextensive rescaling}
\label{s:dynamics}

From the outset, the concept of extensivity played a major role in this paper, and an $N$-dependent rescaling factor was used to enforce extensive energies or Hamiltonians. While this concept turned out to be useful, the discussion in Sec.~\ref{s:interpretation} also pointed out that a more general idea lies behind the $N$-dependent rescaling, namely that of finding the correct scaling such that a certain feature of interest, like the peak of the specific heat in Sec.~\ref{s:interpretation}, is captured asymptotically for large $N$ by a meaningful, nontrivial thermodynamic limit. While enforcing extensivity in many cases achieves this goal, this is not always the case. The following example of the nonequilibrium dynamics of a quantum spin system evolving unitarily under long-range interactions will illustrate a situation where a meaningful and nontrivial thermodynamic limit is obtained with an $N$-dependent scaling factor that does not enforce extensivity of the Hamiltonian, but requires a different power of $N$.

The model considered here is a long-range version of the quantum Ising chain, characterized by the Hamiltonian
\begin{equation}\label{e:LRQuantumIsing}
H_\alpha=-\sum_{i<j}\frac{\sigma^z_i\sigma^z_j}{|i-j|^\alpha},
\end{equation}
where $\sigma^z_i$ denotes the $z$-component of a Pauli operator assigned to lattice site $i$. We assume periodic boundary conditions. The unitary time evolution generated by this Hamiltonian is in general hard to analyze analytically. However, for fully $x$-polarized initial states, time-evolved expectation values of spin operators can be calculated at arbitrary times $t$, e.g.,
\begin{equation}\label{e:sigmaxt}
\braket{\sigma^x_i}(t)=\prod_{j=1}^{N/2} \cos^2\left(\frac{2t}{j^\alpha}\right)
\end{equation}
for even $N$; see Refs.~\onlinecite{Emch66,Kastner11,vdWorm_etal13} for details, derivations, and generalizations.

\begin{figure}\center
\includegraphics[width=0.72\linewidth]{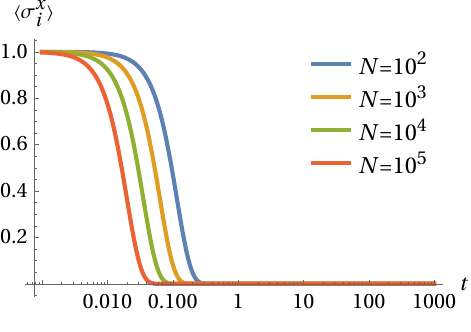}
\includegraphics[width=0.72\linewidth]{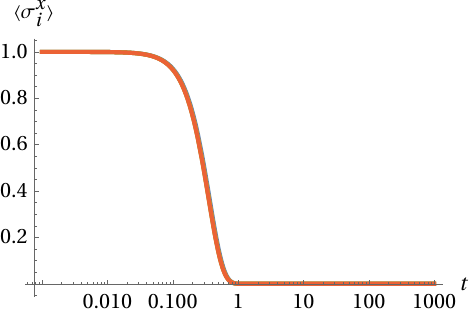}
\includegraphics[width=0.72\linewidth]{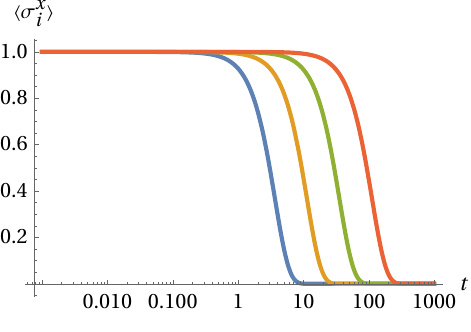}
\caption{\label{f:sigmaxt}
(Color online) The expectation value \eqref{e:sigmaxt} of the spin operator $\sigma_i^x$ at an arbitrary lattice site $i$ for the long-range exponent $\alpha=1/4$ and different system sizes $N$. Time evolution starts from a fully $x$-polarized initial state and is shown on a logarithmic scale. Top: For the Hamiltonian \eqref{e:LRQuantumIsing} without $N$-dependent scaling, the shoulder of the curve is shifted to earlier times with increasing $N$. Center: For the Hamiltonian \eqref{e:LRQuantumIsingCollapse} with an $N$-dependent scaling factor that is different from the one that makes the Hamiltonian extensive. The curves for different system sizes lie on top of each other and are undistinguishable on the scale of the plot. Bottom: For the Hamiltonian \eqref{e:LRQuantumIsingExtensive} with an $N$-dependent scaling factor, chosen such that the Hamiltonian becomes extensive. The shoulder of the curve is shifted to later times with increasing $N$.
}
\end{figure}

The time evolution of the expectation value \eqref{e:sigmaxt} is shown in Fig.~\ref{f:sigmaxt} (top) for a small value of $\alpha$ and for different system sizes $N$. Starting from the initial value $\braket{\sigma^x_i}=1$, the expectation value decays toward its vanishing equilibrium value at later times. The time $\tau$ at which the decay to equilibrium occurs depends on $N$. From the equal spacings between the different curves on a logarithmic time scale, one can infer that $\tau\propto N^q$ with $q<0$ and hence the equilibration time vanishes in the thermodynamic limit. This behavior shares important features with the specific heat of the unscaled all-to-all Ising model in Fig.~\ref{f:SpecificHeat} (top): The ``feature of interest'' (i.e., the decay of the spin expectation value in the one case, and the peak of the specific heat in the other) is shifted toward a singular point (time $t=0$ in the one case and inverse temperature $\beta=1/T=0$ in the other) with increasing system size $N$.

In the case of the specific heat in Fig.~\ref{f:SpecificHeat} (top), this unwanted behavior was removed by making the all-to-all Ising Hamiltonian extensive. Using the same idea for the dynamics of the quantum Ising model, we enforce extensivity by multiplying the Hamiltonian \eqref{e:LRQuantumIsing} with the $N$-dependent prefactor \eqref{e:Prefactor}, yielding
\begin{equation}\label{e:LRQuantumIsingExtensive}
\tilde{H}_\alpha=-N^{\alpha-1}\sum_{i<j}\frac{\sigma^z_i\sigma^z_j}{|i-j|^\alpha}
\end{equation}
for $0\leq\alpha<1$. Computing the time evolution of $\braket{\sigma^x_i}$ generated by the scaled Hamiltonian \eqref{e:LRQuantumIsingExtensive}, which corresponds to the replacement $t\to tN^{\alpha-1}$ on the right-hand side of Eq.\ \eqref{e:sigmaxt}, we find the behavior illustrated in Fig.~\ref{f:sigmaxt} (bottom): The decay to equilibrium now occurs at a time $\tau$ that {\em increases}\/ with increasing system size $N$ like $\tau\propto N^{\tilde{q}}$ with $\tilde{q}>0$. As a result, the equilibration time is pushed to infinity in the thermodynamic limit and the system is stuck in its initial state indefinitely. No interesting physics about the approach to equilibrium of large but finite systems can be deduced from such a constant (in time) thermodynamic limit result.

It turns out that a different $N$-dependent scaling factor, and hence a rescaled but nonextensive Hamiltonian, leads to the desired behavior.\cite{BachelardKastner13,KastnerVdWorm15} Computing the time evolution of $\braket{\sigma^x_i}$ for the Hamiltonian
\begin{equation}\label{e:LRQuantumIsingCollapse}
\bar{H}_\alpha=-N^{\alpha-1/2}\sum_{i<j}\frac{\sigma^z_i\sigma^z_j}{|i-j|^\alpha}
\end{equation}
for $0\leq\alpha<1/2$, we find that the curves for different system sizes $N$ lie nicely on top of each other; see Fig.~\ref{f:sigmaxt} (center). Using the scaled Hamiltonian \eqref{e:LRQuantumIsingCollapse}, a thermodynamic limit result for the time evolution of $\braket{\sigma^x_i}$ would likewise look like the curves in that plot, and hence provides a useful idealization of the decay to equilibrium for large finite systems.

Unlike in equilibrium physics, where extensivity of the Hamiltonian turns out to be the suitable rescaling goal in virtually all circumstances, there is more variability in nonequilibrium dynamics. Not only can different Hamiltonians induce dynamics on different timescales, and hence possibly require other powers of $N$ for rescaling, but different observables or initial states may have similar effects; see Ref.~\onlinecite{vdWorm_etal13} for an example. Classical Hamiltonian dynamics, being governed by second-order differential equations, may require yet another $N$ scaling, as for example in Ref.~\onlinecite{Yamaguchi_etal04}.


\section{Conclusions}
\label{s:conclusions}

Introducing {\em ad hoc}\/ an $N$-dependent scaling factor in the Hamiltonian of a many-body Hamiltonian, as is frequently done when dealing with long-range interacting systems, may look like an unjustified tampering at first sight. We have seen in this paper that introducing such a scaling factor not only serves a purpose, but also has a meaningful physical interpretation and is in many cases reversible. The main messages of the present paper can be summarized as follows:
\begin{itemize}
\item Sections \ref{s:Introduction}--\ref{s:continuum}: The Hamiltonian of a long-range interacting system can be made extensive by multiplying nonextensive terms in the Hamiltonian with a suitable $N$-dependent prefactor.
\item Section \ref{s:finiteN}: Such a rescaling of the Hamiltonian does not constitute an objectionable tampering with the system under investigation but is, for all finite system sizes, a reversible procedure that can be undone by a suitable rescaling of parameters.
\item Sections \ref{s:why} and \ref{s:interpretation}: When taking the thermodynamic limit, however, it does make a significant difference whether or not a scaling factor is used. Without a scaling factor, or with an unsuitably chosen one, a quantity of interest may not converge to a finite limiting value at all; or it may converge, but to a result that does not capture the features of interest of large but finite systems that one intends to study.
\item Section \ref{s:dynamics}: Rescaling the Hamiltonian such that extensivity is enforced leads to meaningful thermodynamic limit results in many cases, but there are cases where a scaled, but nonextensive Hamiltonian is needed.
\end{itemize}

Why does enforcing extensivity often, but not always, lead to the correct scaling? In many cases, an extensive ``reference point'' is built into either the system or the physical formalism. This can be an extensive term in the Hamiltonian, like the magnetic field term in the all-to-all Ising Hamiltonian \eqref{e:CWIsing} or the kinetic energy term in the Hamiltonian \eqref{e:GpotReg} of gravitating point masses, that coexists with the long-range pair potential. But even in the absence of such an explicitly extensive term, extensivity enters indirectly into the formalism of statistical mechanics: Due to the product structure of many-body phase spaces or Hilbert spaces, partition functions grow exponentially with the system size $N$, which leads to an extensive entropic contribution to the free energy or other thermodynamic potentials. It is for this reason that the extensive all-to-all Ising Hamiltonian \eqref{e:CWIsingRescaled_h0} turned out to be suitable for computing a meaningful thermodynamic limit. When computing the decay to equilibrium of the spin expectation value in Sec.~\ref{s:dynamics} on the other hand, there was neither an extensive term in the Hamiltonian in competition with the nonextensive one, nor did we use a physical formalism that brought in extensivity through the backdoor. In the absence of an extensive point of reference, different scaling procedures may then indeed turn out to be suitable.

Finally, it is worth pointing out that long-range interacting systems are not the only class of systems where nonextensivity may occur. For example, the density of states in a strongly constrained system may grow subexponentially, instead of exponentially, with the system size.\cite{WebsterKastner18} The microcanonical entropy will then scale sublinearly with $N$, and an $N$-dependent rescaling of the Hamiltonian is necessary to retain competition between energetic and entropic effects in the thermodynamic limit.

\begin{acknowledgments}
Over the years, the author has benefited from many useful and insightful discussions with colleagues on the topic of this paper. Special thanks go to Alessandro Campa, Nicolò Defenu, Shamik Gupta, Michael Kiessling, Stefano Ruffo, and Hugo Touchette. This research was supported in part by grant NSF PHY-2309135 to the Kavli Institute for Theoretical Physics (KITP).
\end{acknowledgments}

\appendix


\section{Historical remarks}
\label{s:history}
The $1/N$ prefactor in the gravitational potential \eqref{e:GpotAlpha}, or more generally the prefactor $\mathcal{N}_N$ in a long-range lattice model like \eqref{e:LRIsing}, is frequently referred to as ``Kac prescription'' in the literature,\cite{CampaDauxoisFanelliRuffo,CamDauxRuf09,Defenu_etalRMP23} occasionally augmented with a reference to Kac {\em et al.}\cite{KacUhlenbeckHemmer63} The content of that paper, however, is quite different from what is discussed in the present article: Ref.~\onlinecite{KacUhlenbeckHemmer63} deals with exponentially (not algebraically) decaying potentials, and considers a limit that comprises not only a scaling of the strength of the pair interactions, but simultaneously of their {\em range}. To the best of my knowledge, this type of limit goes back to Baker.\cite{Baker61} The authors of Ref.~\onlinecite{KacUhlenbeckHemmer63} refer to the limit as the {\em van der Waals limit} as, for their model of a one-dimensional fluid, it reproduces the equation of state of the van der Waals theory of phase transitions.

While it may be argued that the $1/N$ prefactor in the pair potential of all-to-all connected spin models was in principle already contained in the work of Pierre Weiss\cite{Weiss07} on the mean-field theory of magnetism, it was not written down explicitly. Also, in that reference the $1/N$ prefactor is not so much part of the specifications of a long-range model, but the outcome of the application of a mean-field approximation to a spin system with short-range interactions. Apparently it took until around the 1960s, primarily in the mathematical physics literature as part of the quest to come up with exactly solvable models that undergo a phase transitions, for all-to-all connected lattice models to be considered as models in their own right, with Hamiltonians that contain a $1/N$ prefactor; see Ref.~\onlinecite{EmchDG} and earlier papers by the same author. Going beyond all-to-all interactions, lattice models with power law interactions and the corresponding scaling factors $\mathcal{N}_N$ defined in Eq.~\eqref{e:curlyN} were introduced around the 1990s, when the unconventional statistical physical properties of long-range interacting systems started to receive increased attention.\cite{AnteneodoTsallis98}


\bibliography{../../MK.bib}

\end{document}